      [1999/12/01 v1.4c Il Nuovo Cimento]
\documentclass[rivista]{cimento}

\usepackage{graphicx}
\usepackage[center,footnotesize,hang]{subfigure}

\newcommand{\beq}{\begin{equation}}
\newcommand{\eeq}{\end{equation}}

\title{Tri-bimaximal and bimaximal neutrino mixing from discrete symmetries}
\author{L.~Merlo\from{ins:x}\thanks{e-mail address: merlo@pd.infn.it}}
\instlist{\inst{ins:x} Dipartimento di Fisica `G.~Galilei', Universit\`a di Padova \&
INFN, Sezione di Padova, Via Marzolo~8, I-35131 Padua, Italy}
\PACSes{
\PACSit{11.30.Hv}{Flavour symmetries}
\PACSit{14.60.Pq}{Neutrino mass and mixing}}
\begin{document}

\maketitle

\begin{abstract}
The neutrino oscillation data are well explained by the tri-bimaximal pattern. Recently it has been shown that also the bimaximal pattern could be a very good starting point in order to describe the lepton mixing. In this paper I review both the flavour structures and then I present an explicit example.
\end{abstract}


\section{The lepton mixing matrix}
The solar and the atmospheric neutrino anomalies are well explained by the oscillation of three massive neutrinos: in table \ref{table:OscillationData} there are the results of two independent global fits from \cite{Fogli_Data} and \cite{Schwetz_Data} of the experimental data.

\begin{table}[ht!]
\caption{Neutrino oscillation parameters from independent global fits \cite{Fogli_Data, Schwetz_Data}.}
\label{table:OscillationData}
\begin{center}
\begin{tabular}{lcccc}
\hline
& \multicolumn{2}{c}{Ref.~\cite{Fogli_Data}} & \multicolumn{2}{c}{Ref.~\cite{Schwetz_Data}}\\
parameter & best fit(\@$1\sigma$) & 3$\sigma$-interval & best fit(\@$1\sigma)$ & 3$\sigma$-interval\\
\hline
$\Delta m^2_{21}\:[10^{-5}\mathrm{eV}^2]$
        & $7.67^{+0.16}_{-0.19}$ & $7.14-8.19$
        & $7.65^{+0.23}_{-0.20}$ & $7.05-8.34$\\[1mm]
$|\Delta m^2_{31}|\: [10^{-3}\mathrm{eV}^2]$
        & $2.39^{+0.11}_{-0.8}$ & $2.06-2.81$
        & $2.40^{+0.12}_{-0.11}$ & $2.07-2.75$\\[1mm]
$\sin^2\theta_{12}$
        & $0.312^{+0.019}_{-0.018}$ & 0.26-0.37
        & $0.304^{+0.022}_{-0.016}$ & 0.25-0.37\\[1mm]
$\sin^2\theta_{23}$
        & $0.466^{+0.073}_{-0.058}$ & 0.331-0.644
        & $0.50^{+0.07}_{-0.06}$ & 0.36-0.67\\[1mm]
$\sin^2\theta_{13}$
        & $0.016^{+0.010}_{-0.010}$ & $\leq$ 0.046
        & $0.010^{+0.016}_{-0.011}$ & $\leq$ 0.056\\
\hline
\end{tabular}
\end{center}
\vspace{-1cm}
\end{table}

The pattern of the mixings is characterized by two large angles and a small one: the atmospheric angle $\theta_{23}$ is compatible with a maximal value, but the accuracy admits relatively large deviations; the solar angle $\theta_{12}$ is large, but about $5\sigma$ errors far from the maximal value; the reactor angle $\theta_{13}$ only has an upper bound, which can be parameterized at $3\sigma$ as $\sin^2\theta_{13}\leq\lambda^2_C$, where $\lambda_C\simeq0.23$ is the Cabibbo angle. Measuring the reactor angle in future experiments represents a fundamental task in order to understand the pattern of the neutrino mixing \cite{Reactor}.\\

From the theoretical point of view, for long time, people tried to reproduce models with a maximal angle and a vanishing one. The most general mass matrix which provides $\theta_{13}=0$ and $\theta_{23}$ maximal owns the $\mu-\tau$ symmetry, for which $(m_\nu)_{2,2}=(m_\nu)_{3,3}$ and $(m_\nu)_{1,2}=(m_\nu)_{1,3}$, and it is given by
\vspace{-0.1cm}
\beq
m_\nu=\left(
  \begin{array}{ccc}
    x & y & y \\
    y & z & w \\
    y & w & z \\
  \end{array}
\right)\;.
\vspace{-0.1cm}
\eeq
It depends on four real parameters: the three masses and the solar angle, which can be written in terms of the parameters of the matrix as
\vspace{-0.1cm}
\beq
\sin^2{2\theta_{12}}=\frac{8y^2}{(x-y-z)^2+8y^2}\;.
\label{eq:solar}
\vspace{-0.1cm}
\eeq
Many models have been constructed introducing a flavour symmetry in addition to the gauge group of the Standard Model in order to provide the $\mu-\tau$ symmetric pattern for the neutrino mass matrix, but in all of these the solar angle remains undetermined. It is therefore needed some new ingredient other than the $\mu-\tau$ symmetry to describe correctly the neutrino mixings from the theoretical point of view.\\
A second relevant aspect concerns the maximality of the atmospheric angle: it is a very well known result \cite{AF_extra} that a maximal atmospheric angle can be achieved only when the flavour symmetry is broken. As a consequence, the new flavour symmetry needed to correctly reproduce the three mixing angles has to be broken, spontaneously or softly.

\subsection{The Tri-bimaximal mixing} In what follows I review two relevant flavour patterns, which can be considered an upgrade of the $\mu-\tau$ symmetry. The first one is the tri-bimaximal (TB) pattern \cite{TB} which assumes a vanishing reactor angle, a maximal atmospheric angle and $\sin^2{\theta_{12}}=1/3$. From eq.(\ref{eq:solar}) it results $w=x+y-z$ and therefore the most generic mass matrix of the TB-type is given by
\vspace{-0.1cm}
\beq
m_\nu^{TB}=\left(
  \begin{array}{ccc}
    x & y & y \\
    y & z & x+y-z \\
    y & x+y-z & z \\
  \end{array}
\right)\;.
\vspace{-0.1cm}
\eeq
This matrix satisfies the $\mu-\tau$ symmetry and the so-called magic symmetry, for which $(m_\nu^{TB})_{1,1}=(m_\nu^{TB})_{2,2}+(m_\nu^{TB})_{2,3}-(m_\nu^{TB})_{1,3}$. It is diagonalized by a unitary transformation in such a way that $(m_\nu^{TB})_{\mathit{diag}}=(U^{TB})^T m_\nu^{TB}U^{TB}$, where the unitary matrix is given by
\vspace{-0.1cm}
\beq
U^{TB}=\left(
         \begin{array}{ccc}
           \sqrt{2/3} & 1/\sqrt3 & 0 \\
           -1/\sqrt6 & 1/\sqrt3 & -1/\sqrt2 \\
           -1/\sqrt6 & 1/\sqrt3 & +1/\sqrt2 \\
         \end{array}
       \right)\;.
\vspace{-0.1cm}
\eeq
Note that $U^{TB}$ does not depend on the mass eigenvalues, in contrast with the quark sector, where the entries of the CKM matrix can be written in terms of the ratio of the quark masses. Moreover it is a completely real matrix, since the factors with the Dirac phase vanish (the Majorana phases can be factorized outside).\\
Many flavour symmetries have been used to recover the TB pattern: the most known are the discrete groups $A_4$ \cite{AF_extra,A4other}, $T'$ \cite{FHLM_Tp}, $S_4$ \cite{BMM_S4} and some continuous groups \cite{Continuous}. In all these models, when the symmetry is broken, some corrections to the mixing angles are introduced: in general all of them are of the order of $\lambda_C^2$ and therefore these models suggest a value for the reactor angle which is very small, at most of the order of $0.05$.

\subsection{The bimaximal mixing}
In agreement with the data in table \ref{table:OscillationData}, the reactor angle is suggested to be non-vanishing and, if a value close to the present upper bound is found in the future experiments, this could be interpreted as an indication that the agreement with the TB mixing is only accidental. Looking for an alternative leading principle, it is interesting to note that the data suggest a numerical relationship between the lepton and the quark sectors, known as the complementarity relation, for which $\theta_{12}+\lambda_C\simeq\pi/4$. However, there is no compelling model which manages to get this nice feature, without parameter fixing. Our proposal is to relax this relationship. Noting that $\sqrt{m_\mu/m_\tau}\simeq\lambda_C$, we can write the following expression, which we call weak complementarity relation
\vspace{-0.1cm}
\beq
\theta_{12}\simeq\frac{\pi}{4}-\mathcal{O}\left(\sqrt{\frac{m_\mu}{m_\tau}}\right)\;.
\vspace{-0.1cm}
\eeq
The idea is first to get a maximal value both for the solar and the atmospheric angles and then to correct $\theta_{12}$ with relatively large terms. To reach this task, the bimaximal (BM) pattern can be extremely useful: it corresponds to the requirement that $\theta_{13}=0$ and $\theta_{23}=\theta_{12}=\pi/4$. For the solar angle we can alternatively write that $\sin^2{\theta_{12}}=1/2$, which brings to $w=x-z$ from eq.(\ref{eq:solar}). The most general mass matrix of the BM-type and its diagonalizing unitary matrix are then given by
\vspace{-0.1cm}
\beq
m_\nu^{BM}=\left(
  \begin{array}{ccc}
    x & y & y \\
    y & z & x-z \\
    y & x-z & z \\
  \end{array}
\right)\;,\qquad\quad
U^{BM}=\left(
         \begin{array}{ccc}
           1/\sqrt2 & -1/\sqrt2 & 0 \\
           1/2 & 1/2 & -1/\sqrt2 \\
           1/2 & 1/2 & +1/\sqrt2 \\
         \end{array}
       \right)\;,
\vspace{-0.1cm}
\eeq
where $m_\nu^{BM}$ satisfies to the $\mu-\tau$ symmetry and to an additional symmetry for which $(m_\nu^{BM})_{1,1}=(m_\nu^{BM})_{2,2}+(m_\nu^{BM})_{2,3}$. Note that $U^{BM}$ does not depend on the mass eigenvalues and is completely real, like in the TB pattern.\\
Starting from the BM scheme, the corrections introduced from the symmetry breaking must have a precise pattern: $\delta\sin^2\theta_{12}\simeq\lambda_C$, while $\delta\sin^2\theta_{23}\leq\lambda_C^2$ and $\delta\sin\theta_{13}\leq\lambda_C$ in order to be in agreement with the experimental data. This feature is not trivially achievable.


\section{The model building}
In this part I present a flavour model in which the neutrino mixing matrix at the leading order (LO) is the BM scheme, while the charged lepton mass matrix is diagonal with hierarchical entries; moreover the model allows for corrections that bring the mixing angles in agreement with the data (for details see \cite{AFM_Bimax}). The strategy is to use the flavour group $G_f=S_4\times Z_4\times U(1)_{FN}$, where $S_4$ is the group of the permutations of four objects, and let the SM fields transform non-trivially under $G_f$; moreover some new fields, the flavons, are introduced which are scalars under the SM gauge symmetry, but transform under $G_f$; these flavons, getting non-vanishing vacuum expectation values (VEVs), spontaneously break the symmetry in such a way that two subgroups are preserved, $G_\ell=Z_4$ in the charged lepton sector and $G_\nu=Z_2\times Z_2$ in the neutrino sector. It is this breaking chain which assures that the LO neutrino mixing matrix is the BM pattern and that the charged lepton mass matrix is diagonal and hierarchical. As a consequence the LO lepton mixing matrix $U\equiv (U_\ell)^\dag U_\nu$, the only which can be observable, coincides with $U^{BM}$.\\
Thanks to the particular VEV alignment, the next-to-leading order (NLO) corrections, coming from the higher order terms, are not democratic and the corrected mixings are
\vspace{-0.1cm}
\beq
\sin^2\theta_{12}=\frac{1}{2}-\frac{1}{\sqrt2}(a+b)\;,\qquad
\sin^2\theta_{23}=\frac{1}{2}\;,\qquad
\sin\theta_{13}=\frac{1}{\sqrt2}(a-b)\;,
\vspace{-0.1cm}
\eeq
where $a$ and $b$ parameterize the VEVs of the flavons. When $a$ and $b$ are of the order of the Cabibbo angle, then the $\theta_{12}$ is brought in agreement with the experimental data; in the meantime the reactor angle is corrected of the same amount, suggesting a value for $\theta_{13}$ close to its present upper bound. Note that the atmospheric angle remains uncorrected at this order.\\
\\
It is then interesting to verify the agreement of the model with other sectors of the neutrino physics, such as the $0\nu2\beta$-decay and the leptogenesis. The result of the analysis is that the model presents a normal ordered -- moderate hierarchical or quasi degenerate -- spectrum with a suggested lower bound for the lightest neutrino mass and for the effective $0\nu2\beta$-mass parameter $|m_{ee}|$ of about 0.1 meV. On the other hand it is compatible with the constraints from leptogenesis as an explanation of the baryon asymmetry in the Universe.


\acknowledgments
I thank the organizers of \emph{IFAE 2009 -- Incontri di Fisica delle Alte Energie} for giving the opportunity to present my talk and for the kind hospitality in Bari.
Alike I thank Guido Altarelli and Ferruccio Feruglio for the pleasant and advantageous collaboration.

\end{document}